\documentclass[usenatbib]{mn2e}

\usepackage{ulem}
\usepackage[usenames]{color}
 
\usepackage{graphicx}
\newcommand{\bA}{\bmath{A}}

\newcommand{\bI}{\bmath{I}}

\newcommand{\br}{\bmath{r}}

\newcommand{\bP}{\bmath{P}}

\newcommand{\fracb}[2]{\left(\frac{#1}{#2}\right)}
\newcommand{\text}[1]{\quad\mbox{#1}\quad}
\newcommand{\sub}[1]{{\mbox{\tiny #1}}}

\newcommand{\oder}[2]{\frac{d #1}{d #2}}
\newcommand{\pder}[2]{\frac{\partial #1}{\partial #2}}

\newcommand{\massint}{k}
\newcommand{\cf}{c_{\rm f}}
\newcommand{\ca}{c_{\rm a}}
\newcommand{\cs}{c_{\rm s}}
\newcommand{\vvartheta}{\theta_{\rm v}}

\newcommand{\apj}{ApJ}

\newcommand{\mnras}{MNRAS}

\newcommand{\aap}{A\&A}

\title[Rarefaction acceleration of magnetized GRB jets]
{Rarefaction acceleration of  
ultra-relativistic magnetized jets in gamma-ray burst sources}
\author[S.~S. Komissarov et al.]
{Serguei~S.~Komissarov,$^{1}$\thanks{E-Mail: serguei@maths.leeds.ac.uk~(SSK);
vlahakis@phys.uoa.gr~(NV); arieh@jets.uchicago.edu~(AK)} 
Nektarios~Vlahakis$^{2}$\footnotemark[1] and 
Arieh~K\"onigl$^{3}$\footnotemark[1] \\
$^{1}$Department of Applied Mathematics, The University of Leeds, Leeds, LS2 9GT\\
$^{2}$Section of Astrophysics, Astronomy and Mechanics, 
Physics Department, University of Athens, 15784 Zografos, Athens, Greece\\
$^{3}$Department of Astronomy and Astrophysics and
Enrico Fermi Institute, University of Chicago, 5640 South Ellis Avenue,\\
\ Chicago, IL 60637, USA}

\begin{document}
\date{Received/Accepted}
\maketitle

\begin{abstract}
When a magnetically dominated super--fast-magnetosonic long/soft
gamma-ray burst (GRB) jet leaves the progenitor star, the external
pressure support will drop and the jet may enter the regime of ballistic
expansion, during which additional magnetic acceleration becomes
ineffective. However, recent numerical simulations by Tchekhovskoy et
al. have suggested that the transition to this regime is accompanied by
a spurt of acceleration. We confirm this finding numerically and
attribute the acceleration to a sideways expansion of the jet,
associated with a strong magnetosonic rarefaction wave that is driven
into the jet when it loses pressure support, which induces a conversion
of magnetic energy into kinetic energy of bulk motion. This mechanism,
which we dub {\it rarefaction acceleration}, can only operate in a
relativistic outflow because in this case the total energy can still be
dominated by the magnetic component even in the super--fast-magnetosonic
regime. We analyse this process using the equations of
relativistic MHD and demonstrate that it is more efficient at converting
internal energy into kinetic energy when the flow is magnetized than in
a purely hydrodynamic outflow, as was found numerically by Mizuno et al.
We show that, just as in the case of the magnetic acceleration of a
collimating jet that is confined by an external pressure distribution --
the {\it collimation acceleration} mechanism --
the rarefaction-acceleration process in a magnetized jet is a
consequence of the fact that the separation between neighbouring
magnetic flux surfaces increases faster than their cylindrical radius.
However, whereas in the case of effective collimation-acceleration the
product of the jet opening angle and its Lorentz factor does not exceed
$\sim 1$, the addition of the rarefaction-acceleration mechanism makes
it possible for this product to become $\gg 1$, in agreement with the
inference from late-time panchromatic breaks in the afterglow light
curves of long/soft GRBs.
\end{abstract}
                                                              

\section{Introduction}
\label{introduction}

In the `standard' model of long-duration, soft-spectrum gamma-ray
bursts (GRBs; e.g.  \citealt{Pir05}), the prompt high-energy emission
arises in ultra-relativistic (bulk Lorentz factor $\Gamma \ga 10^2$),
highly collimated (opening half-angle of a few degrees) jets. The high
Lorentz factors are inferred from the requirement of a sufficiently low
opacity to photon-photon annihilation or to scattering by photon
annihilation-produced electron-positron pairs \citep[e.g.][]{LS01}. 
The high collimation makes it possible to reduce the total flow energy
down to values that are comparable to the energy of stellar explosions. 

\begin{figure*}
\includegraphics[width=85mm]{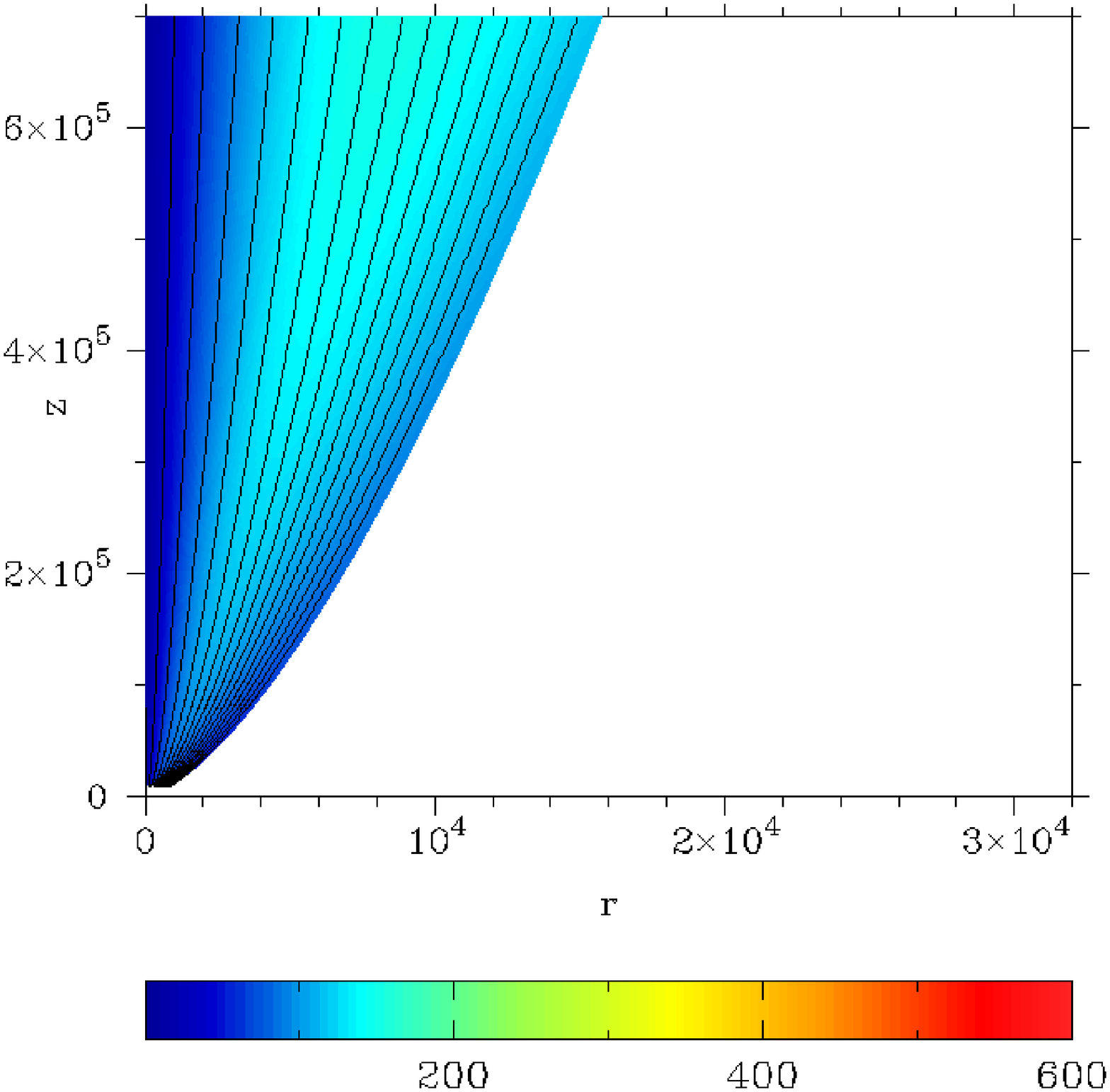}
\includegraphics[width=85mm]{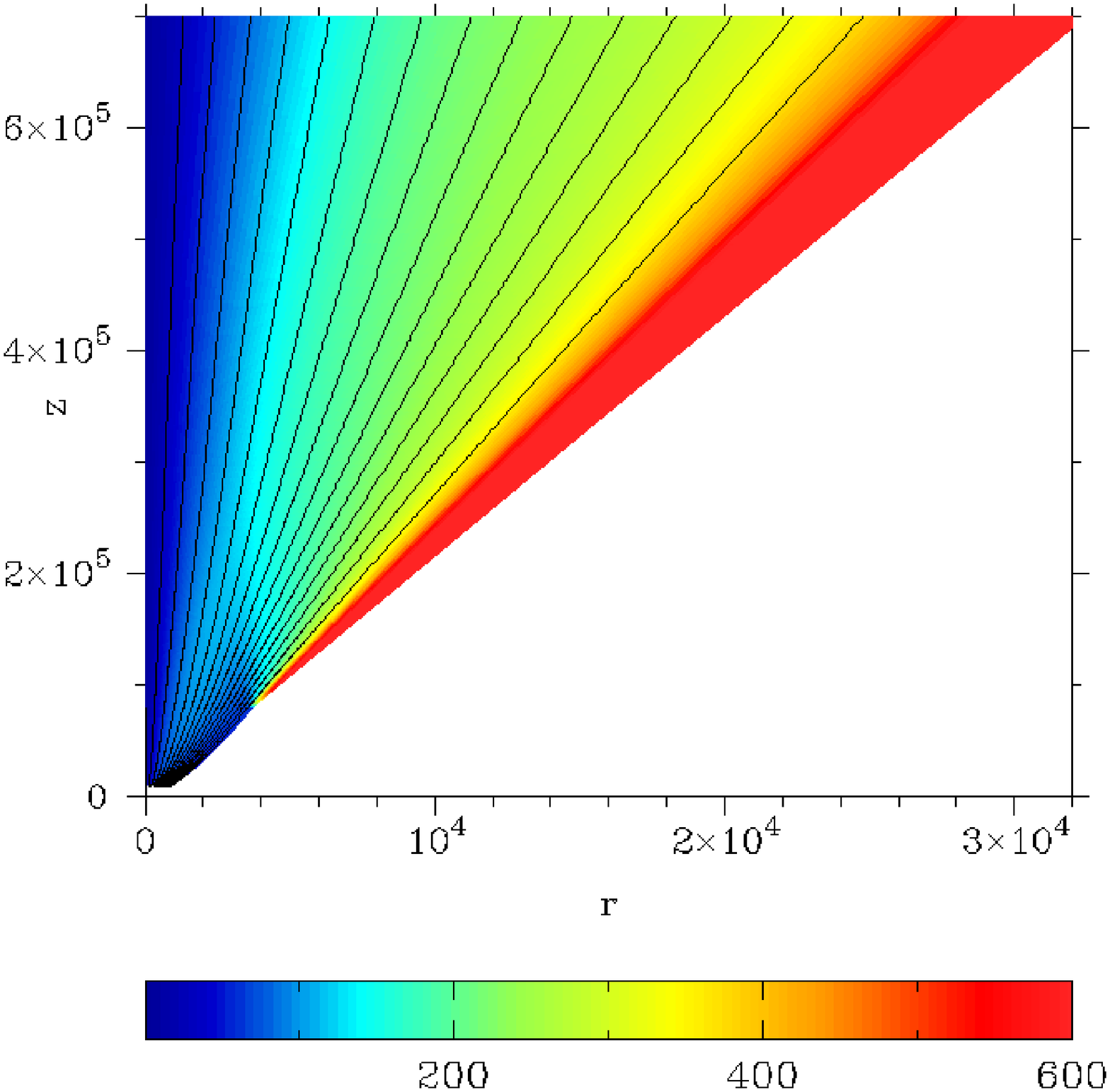}
\caption{Lorentz factor and magnetic field lines  in models B1 
(left panel) and B1b (right panel).}
\label{lorentz}
\end{figure*}

As the jets make their way through the interstellar medium they sweep
the ambient gas into a thin relativistic shell that itself becomes
a strong source of electromagnetic radiation. 
After a sufficiently large amount of gas is swept up,
the shell begins to decelerate. Simplified models of this interaction
predict that a panchromatic break would then occur in 
the afterglow light curve, provided that $\Gamma\theta_{\rm j}\gg1$, where
$\theta_{\rm j}$ is the jet opening half-angle. The detection of such `jet
breaks' on time scales of $\ga 1\,$ day in several (primarily long/soft) GRB
afterglows has been a key reason for the widespread adoption of the jet
model for these sources.
The observed break
parameters have made it possible to deduce the opening
half-angle of these jets. The inferred values are strongly model
dependent, although they usually lie the range of $1^\circ-10^\circ$
\citep[e.g.][]{Rho99,Sari99}.
Although more recent
observations by the {\it Swift}\/ satellite have revealed that late-time
panchromatic jet
breaks are not that common and that various aspects of the jet model may
need to be modified \citep[e.g.][]{Mes06,Pan07,Lia08,Rac09}, these
breaks remain the strongest evidence for collimated outflows in GRB
sources.

The supernova connection of long/soft GRBs provides strong support for
theoretical models of their jet engines that invoke dying massive and
rapidly rotating stars.  These models are generally divided into two
groups depending on the mechanism of jet acceleration. In the first
group the acceleration is driven by the thermodynamic pressure of plasma
heated to ultra-relativistic temperatures via the annihilation of
neutrinos and antineutrinos emitted by the accretion disk formed around
the central black hole, thus tapping the disk thermal energy. In the
second group the acceleration is driven by magnetic stresses,
tapping the rotational energy of the disk or the central compact object
(neutron star or black hole). At present, both the magnetic and the
thermal mechanisms seem possible, although the lack of detection of a
thermal component in the spectra of some GRBs 
is consistent with the notion that, at least in certain cases, the
outflow is initially magnetically dominated \citep{ZP09}.

The magnetic jet acceleration mechanism has been the subject of
theoretical study for many years. Due to the mathematical complexity of
magnetohydrodynamics (MHD), which is even more pronounced in the
relativistic limit, it has been possible to find analytical and
semi-analytical solutions only for a rather limited number of problems
characterized by a high degree of symmetry.  In fact, there is only one
available {\it exact}\footnote{ {\it Exact} in the sense that the full
system of MHD equations -- including the fluid-inertia terms and the
trans-field component of the momentum equation -- is integrated.}
solution of the relativistic MHD equations including thermal and
magnetic effects, the self-similar model of \cite{VK03a}. The advance of
numerical methods for relativistic MHD during the last decade has opened
a new direction of study that has already resulted in significant
progress
\citep[e.g.][]{K01,K04,M06,KB07,KVK07,T08,B08,KVK09,T09a,KB09,T09b,MB09,B09}.

\begin{figure*}
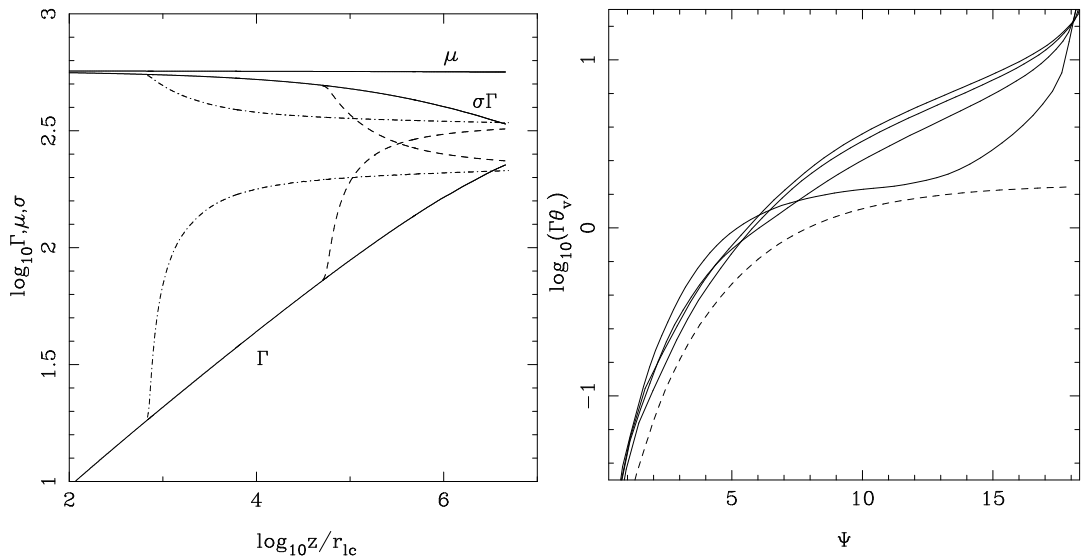

\includegraphics[width=73mm,angle=-90]{figures/sigma2.eps}
\includegraphics[width=72mm,angle=-90]{figures/lortheta1.eps}
\caption{Left panel: Evolution of $\Gamma$, $\sigma\Gamma=\mu-\Gamma$ and 
$\mu$ along the magnetic surface enclosing $~80\%$ of the total 
magnetic flux ($\Psi=15$) in models B1 (solid line), B1a (dash-dotted line), 
and B1b (dashed line), with the distance measured in units of the
light-cylinder radius $r_{\rm lc}$ of the source. 
Right panel: Evolution of $\Gamma\vvartheta$, where $\vvartheta$ is the
opening half-angle,
along the jet in model B1b. The solid lines show the variation of this
parameter across the jet as a function of 
the poloidal  magnetic flux $\Psi$
for $z=1,2,4$ and $8\times10^5\, r_{\rm lc}$, with higher values of $z$
corresponding to  higher curves at $\Psi=15$.
The dashed curve shows
$\Gamma\vvartheta$ at $z=8\times10^5\, r_{\rm lc}$ in model B1.}
\label{sigma}
\end{figure*}

In particular, \citet{KVK09} investigated the magnetic acceleration of
ultra-relativistic flows within channels of prescribed geometry,
$z\propto r^a$ (where $r$ and $z$ are dimensionless cylindrical
coordinates) , determined by the shape of coordinate surfaces of
elliptical coordinates. Such shapes correspond to power-law
distributions of the confining pressure that can approximate the
expected distributions in the envelopes of the progenitor stars in both
the collapsar and the magnetar scenarios (which correspond to the dying
star leaving behind a black hole or a neutron star, respectively) as
well as the effect of a confining disk wind. Among other results, they
found that in the case of a gradually widening channel, $a<1$, the
acceleration is not efficient, whereas in the case of a channel with
gradually increasing collimation, $a>1$, the acceleration is effective
but the asymptotic flow obeys the inequality $\Gamma \theta_{\rm j}\le
1$. These numerical results have been strengthened by complementary
theoretical analyses \citep{KVK09,L09}, which led to the question of
whether the magnetic models can accommodate the jet breaks at
all. Recently, a similar numerical study was carried out by
\citet{T09b}, who confirmed the findings of \citet{KVK09}. They have,
however, also considered a somewhat modified setup, wherein at some
distance from the origin, roughly corresponding to the stellar surface,
the channel geometry changes from progressively collimating to
progressively de-collimating.  In this setup \citet{T09b} observed a
remarkable change in the jet behaviour: beyond the point where the
geometry changed, the jet speed underwent a strong boost that was
accompanied only by a very small increase in the jet opening angle. As a
result, the asymptotic flow had $\Gamma\theta_{\rm j} \gg 1$, which
made it possible for late-time jet breaks to occur. The exact shape of the
channel above the transition point did not seem to matter.

In this paper we describe (Section~\ref{simulations}) simulations that
confirm the results of \citet{T09b} and then analyse
(Section~\ref{interpretation}) the underlying physical mechanism. We
discuss and summarize our results in Section~\ref{conclusion}.

\section{Numerical simulations}
\label{simulations}

The numerical method is exactly the same as in \citet{KVK09} and we
refer the reader interested in technical details to 
that paper. Here we only remark that our numerical code is
based on the Godunov-type scheme for relativistic MHD \citep{K99} and
that we look for steady-state axisymmetric solutions using
time-dependent simulations with time-independent boundary conditions.

For our purpose we selected model B1 from \citet{KVK09}, for which the
channel shape parameter is $a=3/2$.  This model describes a cold flow
with values of the field-line constant $\mu$, defined as the
energy flux per unit rest-mass energy flux, as large as $\mu=620$. This
constant sets the upper limit on the Lorentz factor that can be achieved
in this model via ideal MHD mechanism.\footnote{ 
In terms of the energy budget, magnetic
acceleration represents the conversion of Poynting flux into kinetic
energy. The ratio of the kinetic energy flux to the rest-mass energy
flux is equal to the bulk Lorentz factor of the flow. Thus, when the
Poynting flux in the jet is fully converted into kinetic energy, the
acceleration process stops and one has $\mu=\Gamma$.} The initial
ratio of the Poynting flux to the hydrodynamic energy flux is $\sigma_0
= \mu/\Gamma_0-1 \simeq \mu$, as the initial Lorentz factor, $\Gamma_0$,
is close to unity, corresponding to a sub-Alfv\'enic flow.  The base
rotation is uniform, with the dimensionless light-cylinder radius
$r_{\rm lc}$ being $\simeq 1.6$; we used the distance of the inlet
boundary from the origin as the unit of length.

To examine the effect of changing the channel shape, we map the solution
at $z\simeq 10^3$ and $7\times10^4$ onto the inlet boundary of a new
grid corresponding to a conical channel of the same local radius and
with a vertex located at the origin. In this way we introduce a 
change in the channel shape to a profile that results in less
collimation. We then proceed with the
simulations on the new grid, following the same procedure as in model
B1. The two solutions obtained in this way, which we denote as B1a and B1b, 
are analysed below.

Fig.~\ref{lorentz} shows the overall geometry
 as well as the shape of the magnetic surfaces and the evolution of the Lorentz 
factor in models B1 and B1b. One can see that, in contrast with the
situation in
 model B1, the field 
lines in model B1b straighten out. One might think that this reflects the conical 
shape of the channel, but this is not so. In fact, the jet is separated from 
the channel wall by 
a near vacuum.\footnote{In the simulations, the mass density 
in the vacuum zone is kept above zero 
by numerical effects.}
This is the reason why the red-coloured boundary layer in the right
panel of Fig.~\ref{lorentz} is free from magnetic field lines. A similar
separation has been seen in model E of \citet{KVK09} and, we believe, 
also in
the simulations of \citet{T09b}.  As a result, the wall and the jet are
causally disconnected and the precise shape of the wall does not matter.

In model B1b the jet Lorentz factor approaches its maximum possible
value $\Gamma_\sub{max}=\mu$ at the jet boundary, signalling a total
conversion of the Poynting flux.  The acceleration is weaker in the jet
interior but, as one can clearly see in Fig.~\ref{lorentz}, it is still
more effective than in model B1.  This is further illustrated in the
left panel of Fig.~\ref{sigma}, which compares the acceleration in both
models along the magnetic surface enclosing $80\%$ of the jet's total
poloidal magnetic flux. 
In the case of model B1a, where the channel opens up
much earlier, the jet also passes through a phase of rapid
acceleration. However, the acceleration slows down dramatically when
the jet enters the phase of ballistic expansion. As a result, the final
Lorentz factor in this model is even lower than in model
B1. As one can see from Fig.~\ref{lorentz}, the opening 
angle of the jet does
not change much after passing the 
point where the channel widens, in agreement
with the results of \citet{T09b}. Thus, the product $\Gamma
\vvartheta$, where $\vvartheta$ is the local opening half-angle, 
is expected to increase following the rapid increase of the Lorentz factor.  
This is indeed the case, as one can see in the right panel of Fig.~\ref{sigma}.
In model B1b this product is much larger than in model B1, approaching
values that are $\gg 1$ near the jet boundary. Concluding this section, we 
reiterate that our results are in very good agreement with those
obtained in \citet{T09b}, thus confirming that the effect is real and
must have a robust physical basis.
  
\section{Acceleration mechanism}
\label{interpretation}

\begin{figure}
\includegraphics[width=83mm,angle=-90]{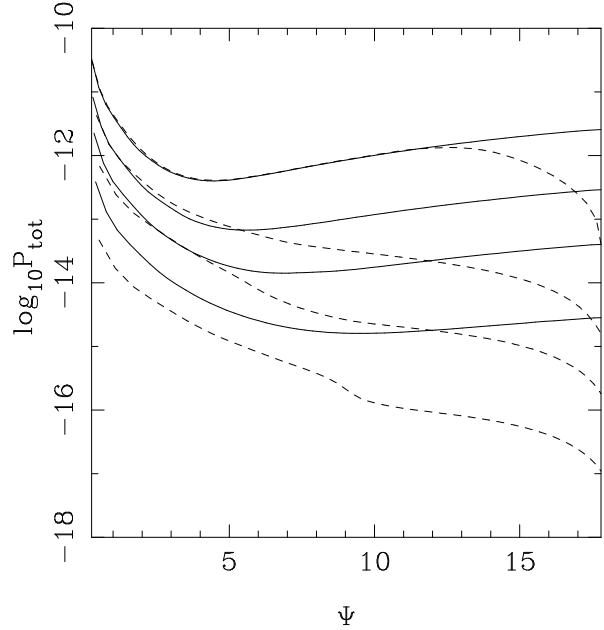}
\caption{Propagation of a rarefaction wave across the jet in model 
B1b. The solid lines show the magnetic pressure distribution 
as a function of the poloidal magnetic flux 
at $z=1,3,9,30\!\times\!10^5\, r_{\rm lc}$ for model B1, whereas the
dashed lines show the corresponding  distributions for model B1b . In
this plot, the higher the value of $z$, the lower the corresponding curve.}
\label{pressure}
\end{figure}

Once the jet separates from the wall in models B1a and B1b it enters the
phase of free expansion and eventually becomes a ballistic conical
outflow with radial streamlines. During the transition from the one
regime to the other a strong fast-magnetosonic rarefaction wave
propagates into the jet. Since at this point the jet is causally
connected \citep[see][]{KVK09}, the wave is not confined to a boundary
layer but propagates all the way to the jet core.  This is clearly seen
in Fig.~\ref{pressure}, which compares the distributions of the magnetic
pressure $b^2/8\pi$ (where $b$ is the magnetic field amplitude in the
comoving frame) in models B1 and B1b downstream from the
channel-widening point. The observed jet acceleration is apparently
related to the properties of this wave.  In fact, this phenomenon has
already been seen in other numerical simulations of both purely
hydrodynamic and magnetized flows \citep{AR06,M08}. Here we elucidate
its physical nature by considering a simpler one-dimensional
problem of relativistic expansion into vacuum in a slab geometry. The
rarefaction wave in this case is described by a self-similar solution
known as a `simple wave'. Although such a flow is not identical to the
one in our simulated jets, it nevertheless captures the underlying
physical mechanism.

\subsection{Simple waves}

Consider a quasilinear hyperbolic system 
\begin{equation}
\pder{\bP}{t} + \bA\pder{\bP}{x} = 0
\label{h-system}
\end{equation}
where $\bP=(P_1,P_2,\dots,P_n)^T$ is a vector of dependent variables,
e.g. the gas pressure, density, etc., and $\bA(\bP)$ is an $(n\times n)$
matrix. When the initial ($t=0$) configuration describes two uniform
states separated by a discontiniuty at $x=0$ (a Riemann problem), the
system allows self-similar solutions that depend only on
$\xi\equiv x/t$. In general, the solution involves a combination of shock
and rarefaction waves, but in the case of expansion into
vacuum it describes a single rarefaction wave (a simple wave).  In terms
of the new independent variable $\xi$, the system of
equations~(\ref{h-system}) reduces to
\begin{equation}
(\bA-\xi\bI)\oder{\bP}{\xi}=0\,
\label{char-eq}
\end{equation}   
where $\bI$ is the unit matrix. 
This equation has the form of the characteristic equation 
for the matrix $\bA$, and thus $\xi$ is equal to one of the eigenvalues
of $\bA$,
\begin{equation}
\xi = \lambda(\bP) \,,
\label{eq-of-char}
\end{equation}
with
\begin{equation}
\oder{\bP}{\xi} = \br(\bP)
\label{sw-eq}
\end{equation}
being the corresponding right eigenvector. 
Each eigenvalue represent the phase speed of a particular wave, 
whereas the right eigenvector determines the connection between the
variations of the dependent variables that are induced by the wave. 
Equation~(\ref{sw-eq}) can be conveniently written as 
\begin{equation}
\frac{dP_1}{r_1}=\frac{dP_2}{r_2}=\dots=\frac{dP_n}{r_n}\, . 
\label{sw-eq1}
\end{equation}
Integrating this system, one finds $P_i=P_i(P_1)$, where $i=2\dots n$,
and then equation~(\ref{eq-of-char}) (the equation of characteristics)
can be used to obtain $P_1=P_1(\xi)$.

The systems of equations for relativistic hydrodynamics and for relativistic 
MHD can be written in the form of equation~(\ref{h-system}). 
The eigenvalues and eigenvectors for all hyperbolic waves can be
found, for example, in 
\citet{A89} and \citet{K99}. Simple waves have been analysed in 
\citet{MM94} for purely hydrodynamic flows and in \citet{RMPIM05} 
for magnetized flows in which the magnetic field is tangential to the
discontinuity and orthogonal to the flow velocity.

\subsubsection{Case 1}
\label{case1}

Consider a simple one-dimensional problem with plane geometry describing
the evolution of an initial discontinuity that separates a uniform
magnetized cold plasma at rest on the left and vacuum on the right.
This problem is related to the transverse expansion of our simulated
jets after they lose external support, as seen in the jet comoving
frame.  The discontinuity decays into a single wave, namely the simple
fast-magnetosonic rarefaction wave.  The left front of this wave (the
head) propagates into the left state with the local fast-magnetosonic
speed and the right front (the tail) moves into vacuum with some finite
terminal speed. In such a simple geometry, the Lorentz force only
involves the magnetic pressure, and the equations of MHD reduce to those
of hydrodynamics of an ideal gas with a polytropic index $\gamma=2$. In
this case the effective sound speed $a$ is equal to the
fast-magnetosonic speed, which satisfies
\begin{equation}
\left(\frac{a}{c}\right)^2=\frac{b^2}{4\pi\rho c^2+b^2} =
\frac{\sigma}{1+\sigma}\,, 
\label{sound}
\end{equation}
where $\rho$ is the rest-mass density and
$\sigma=b^2/4\pi\rho c^2$ is the local magnetization parameter.
This property allows us to utilize the results obtained for the
corresponding problem in relativistic gas dynamics.
In particular, the integration of equation~(\ref{sw-eq1}) yields the integral 
across the rarefaction
\begin{equation}
   J_+=\fracb{1+v_x/c}{1-v_x/c}\fracb{1+a/c}{1-a/c}^2
\end{equation}
\citep{MM94}, where $v_x$ is the local flow speed. 
This equation can be used to deduce $v_0$, the wave's expansion speed
into vacuum. Specifically, by equating the value of the above invariant
as derived from the given left state to the value obtained at the
boundary with the vacuum (where $a=0$), one finds
\begin{equation}
  v_0=\frac{J_+ -1}{J_+ +1}\, c\, ,
\end{equation}
\begin{equation}
\Gamma_0 = (1-v_0^2/c^2)^{-1/2} =\frac{1+a^2/c^2}{1-a^2/c^2} \, .
\label{exp_gamma}
\end{equation}
For example, if in the left state, where $v_x=0$,
we have $b^2=4\pi\rho c^2$, and hence $a=c/\sqrt{2}$,
then the Lorentz factor of expansion into vacuum is $\Gamma_0=3$.
Next, consider another reference frame that moves along the initial
discontinuity with a Lorentz factor $\Gamma_j$ -- this corresponds to
the jet source frame in our simulations. The Lorentz factor of plasma at
the right front of the rarefaction wave in this frame is
\begin{equation}
   \Gamma=\Gamma_j\Gamma_0\, ,
\label{Lor}
\end{equation}
which in our example is three times larger than $\Gamma_j$.
Thus, the seemingly weak acceleration in the jet frame
may correspond to a huge boost in the jet source frame:
in our example the Lorentz factor increases from $\Gamma_j=200$ to
the left of the rarefaction wave to $\Gamma=600$ at the boundary with
vacuum.  This is the essence of the acceleration mechanism that operates
in the free expansion regime of the jet simulations. However,  
the jet expansion problem cannot be
completely reduced to the one considered here since it is inherently two
dimensional (see Section~\ref{bunching} and Appendix~\ref{appB}).
The results can therefore be expected to be {\it quantitatively}\/
different, as one can readily verify. Indeed, along a magnetic
surface of a steady jet the energy flux per unit rest mass flux is
conserved, which for a cold flow leads to the integral 
\begin{eqnarray}
  \mu_j = \Gamma(1+\sigma)  \,,
\nonumber
\end{eqnarray}
where $\mu_j=\Gamma_j(1+\sigma_j)$ and where $\Gamma_j$ and $\sigma_j$
represent the initial jet Lorentz factor and magnetization, respectively.
Thus, the Lorentz factor at the tail of the rarefaction wave, where 
the magnetization has decreased to zero, has to be
$\Gamma=\Gamma_j(1+\sigma_j)$. In contrast, equations~(\ref{exp_gamma})
and~(\ref{Lor}) imply that $\Gamma=\Gamma_j(1+2\sigma_j)$, 
where $\sigma_j$ is the magnetization of the undisturbed left state of
the Riemann problem (which we identify with the initial
magnetization in the steady-jet problem).

\subsubsection{Case 2}
\label{case2}

Here we consider a more complicated Riemann problem where the velocity of 
the left state is not zero but has a component tangent to the discontinuity,  
$\bmath{v}=(0,0,v_j)$. This corresponds to the jet expansion as seen in 
the jet source frame.  The magnetic field of the left state is   
$\bmath{B}=(0,\Gamma_j b_j,0)$, where $b_j$ corresponds to the jet 
azimuthal magnetic field as measured in the comoving frame. In addition, 
we no longer assume that the left state is cold, which will allow us to 
compare the purely hydrodynamic and MHD cases. In the rarefaction wave
$\bmath{v}=(v_x,0,v_z)$ and $\bmath{B}=(0,\Gamma b,0)$. This case has
been analysed in \cite{RMPIM05}. Integration of the simple wave
equations~(\ref{sw-eq1}) leads to the integrals 
\begin{equation}
  I_s=s \,,
\label{Is}
\end{equation}  
\begin{equation}
  I_b=b/\rho \,,
\label{Ib}
\end{equation}  
\begin{equation}
  I_z=h_g u_z
\label{Iz}
\end{equation}
and
\begin{equation}
I_+=\frac{1}{2}\ln\fracb{1+v_x/c}{1-v_x/c}
-\int_\rho^{\rho_j} X(\rho) \frac{d\rho}{\rho}  \ ,
\label{Ip}
\end{equation}  
where $s$ is the specific entropy, $u_z=\Gamma (v_z/c)$ 
is the $z$-component of the 4-velocity, 
\begin{equation}
h_g=\frac{w+b^2/4\pi}{\rho c^2}
\end{equation}  
is the generalized specific enthalpy and 
\begin{equation}
X(\rho)=
\frac{1}{1+u_z^2}
\left[1+u_z^2 \left(1-\frac{\cf^2}{c^2}\right) \right]^{1/2}
\frac{\cf}{c} \ .
\end{equation}  
Here $\cf$ is the fast-magnetosonic speed in the direction normal to the
magnetic field as measured in the fluid frame, with
\begin{equation}
\left(\frac{\cf}{c}\right)^2 = \left(\frac{\ca}{c}\right)^2
+\left(\frac{\cs}{c}\right)^2 - \left(\frac{\ca}{c}\right)^2
\left(\frac{\cs}{c}\right)^2  \,,
\end{equation}  
$\ca$ is the Alfv\'en speed, with
\begin{equation}
\left(\frac{\ca}{c}\right)^2 = \frac{b^2}{b^2+4\pi w} \ ,
\end{equation}  
$\cs$ is the sound speed, with
\begin{equation}
\left(\frac{\cs}{c}\right)^2 = \frac{1}{h} \left(\pder{p}{\rho}\right)_s
\ ,
\end{equation}  
and $h=w/\rho c^2$ is the specific enhalpy. 
The function $X(\rho)$ is a function of density only and
the other variables are eliminated via the integrals~(\ref{Is}),
(\ref{Ib}) and~(\ref{Iz}) for a given equation of state. 
The values of these integrals are dictated by the left state of the problem. 
In particular, 
\begin{eqnarray}
 I_+=0 \text{and} I_z= \mu_j (v_j/c),
\nonumber \end{eqnarray}
where 
\begin{eqnarray}
\mu_j = h_j\Gamma_j(1+\sigma_j) 
\nonumber \end{eqnarray}
and 
\begin{eqnarray}
  \sigma_j=b_j^2/4\pi w_j.
\nonumber \end{eqnarray}
In these expressions we use the index `$j$' to indicate the left state
variables as this state corresponds to the undisturbed jet. Also,
$\sigma_j$ is the magnetization parameter and $\mu_j$ is the energy
integral that incorporates a thermal contribution \citep[see][]{KVK09}.   

In this case the characteristics equation~(\ref{eq-of-char}) yields 
\begin{equation} 
  \frac{\xi-v_x}{1-v_x \xi/c^2} = - \cf 
\left[1+u_z^2 \left(1-\cf^2/c^2\right) \right]^{-1/2}. 
\label{EQ-CHAR1}  
\end{equation}
This result has a straightforward interpretation. 
On the left-hand side one immediately recognizes the 
relativistic expression for the addition of velocities, whereas the right-hand 
side gives the speed of a fast-magnetosonic wave as measured in a frame 
that moves along the $x$ axis with the same speed as the fluid, i.e. $v_x$ 
(see Appendix~\ref{Eq20}).

The procedure for constructing the self-similar solutions using the 
above analytical results is the following. 
>From equations~(\ref{Is})--(\ref{Ip}) and the equation of state,
one finds the functions $w(\rho)$, $b(\rho)$, $u_z(\rho)$, $v_x(\rho)$
and the thermal pressure $p(\rho)$.\footnote{
For a polytropic equation of state with index $\gamma$, the enthalpy is
$w=\rho c^2 + \left[\gamma/(\gamma-1)\right] p$, in which case
equation~(\ref{Is}) becomes $p/\rho^\gamma=\;$const.}
Finally, equation~(\ref{EQ-CHAR1}) allows one to obtain $x/t=\xi(\rho)$. 
In particular we find that the head of the rarefaction wave propagets
with the speed 
\begin{eqnarray}
\xi_{\rm head} = -c_{{\rm f}_j}                
\left[1+u_{z_j}^2 \left(1-c_{{\rm f}_j}^2/c^2\right) \right]^{-1/2},
\nonumber \end{eqnarray}
whereas the wave's tail advances with the speed 
\begin{eqnarray}
 \xi_{\rm tail}=v_x(0)\,. 
\nonumber \end{eqnarray}

Although in general the solutions can only be found numerically, one can
derive fully analytic results in the ultra-relativistic limit, as follows.
From, equation~(\ref{Iz}) one has 
\begin{equation}
   u_z = \mu_j \frac{v_j}{c} \frac{1}{h_g} \,.
\label{uz} 
\end{equation}
For highly relativistic jets $v_z\simeq v_j \simeq c$ and $v_x\ll c$. 
Furthermore, $p/\rho c^2\to 0$ and $b^2/\rho c^2 \to 0$ as $\rho/\rho_j\to 0$.
Equation~(\ref{uz}) then yields
\begin{equation}
  \Gamma \to \mu_j \text{as} \rho/\rho_j \to 0\,. 
\label{boundary-G} 
\end{equation}
The Lorentz factor at the tail of the rarefaction wave (i.e. at the
boundary with vacuum) is thus found to equal the value expected in a
steady jet, as discussed in Section~\ref{case1}. The generalized
one-dimensional Riemann problem considered here (Case 2) therefore
provides a better representation of the simulated two-dimensional jet
problem than the simpler model considered above (Case 1).

\begin{figure*}
\includegraphics[width=43mm,angle=0]{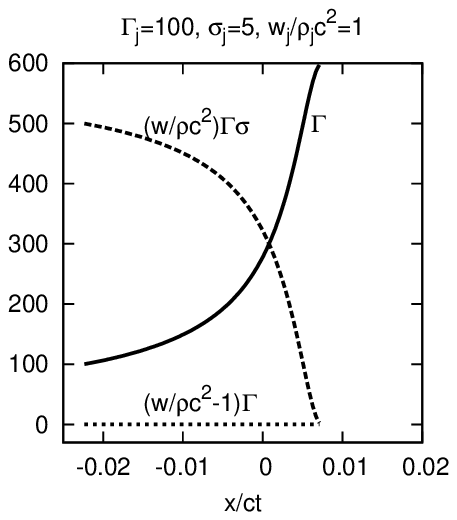}
\includegraphics[width=43mm,angle=0]{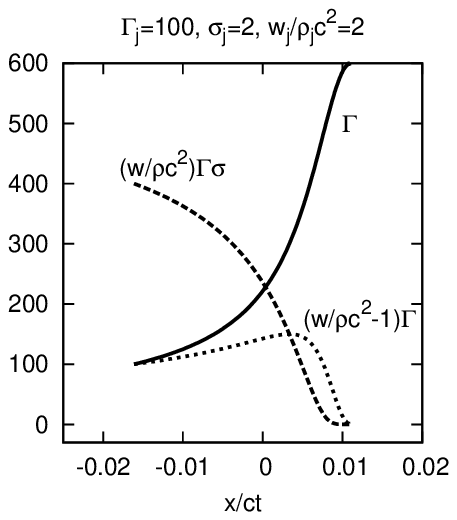}
\includegraphics[width=43mm,angle=0]{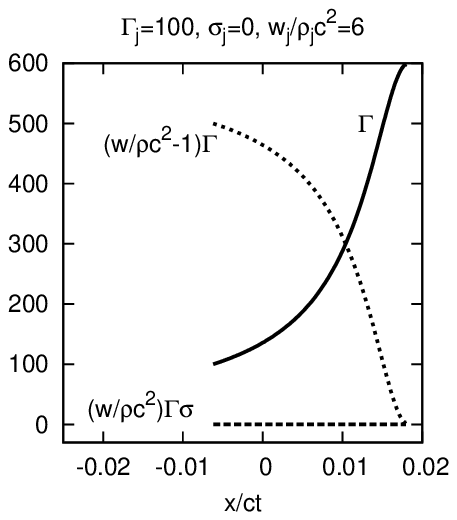}
\includegraphics[width=43mm,angle=0]{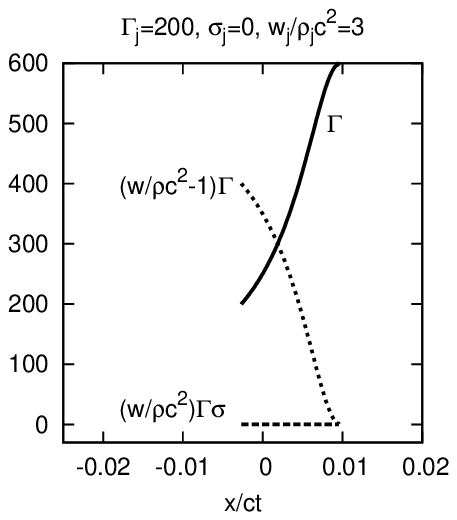}
\\
\includegraphics[width=43mm,angle=0]{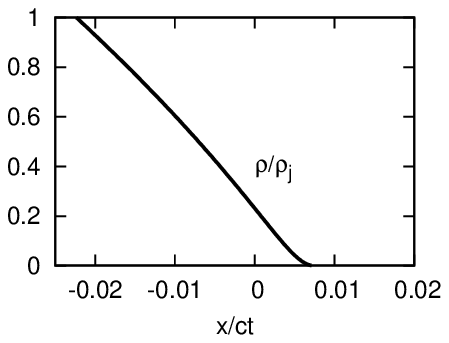}
\includegraphics[width=43mm,angle=0]{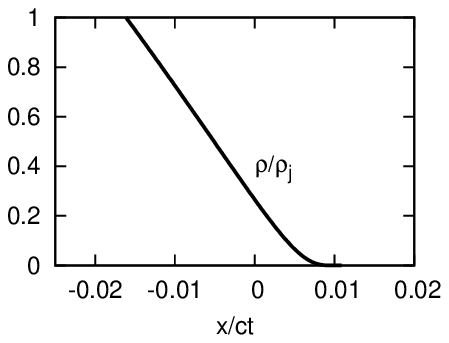}
\includegraphics[width=43mm,angle=0]{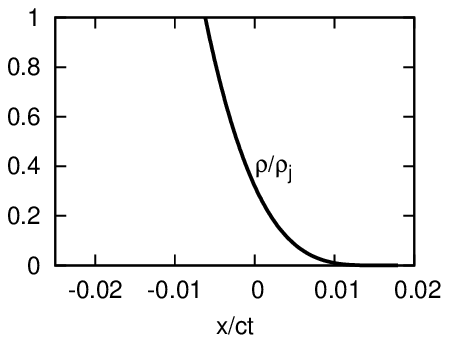}
\includegraphics[width=43mm,angle=0]{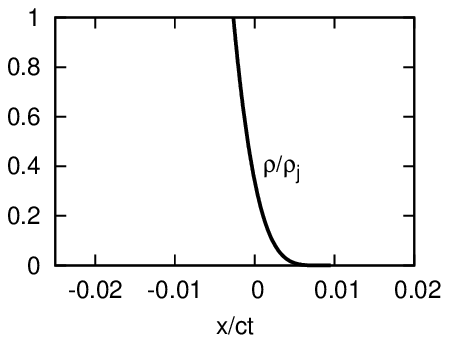}
\\
\includegraphics[width=43mm,angle=0]{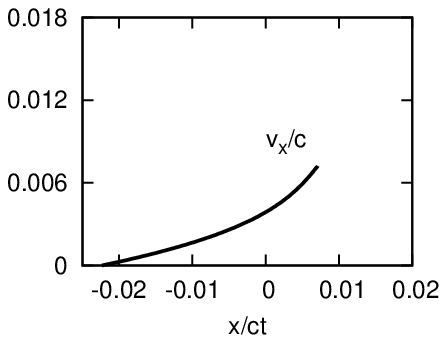}
\includegraphics[width=43mm,angle=0]{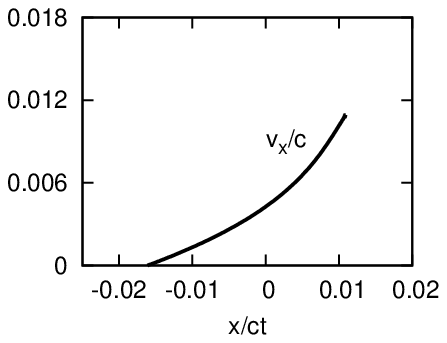}
\includegraphics[width=43mm,angle=0]{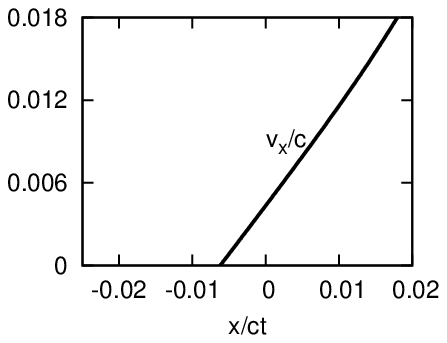}
\includegraphics[width=43mm,angle=0]{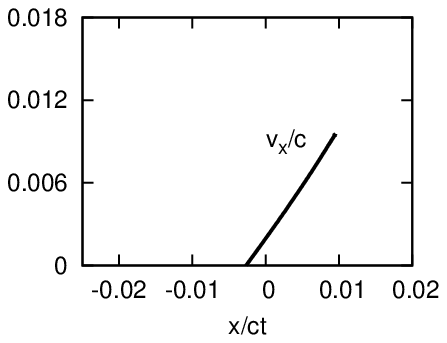}
\\
\includegraphics[width=43mm,angle=0]{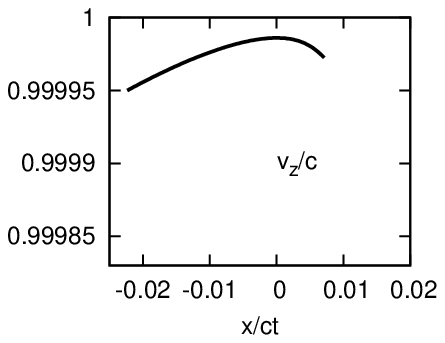}
\includegraphics[width=43mm,angle=0]{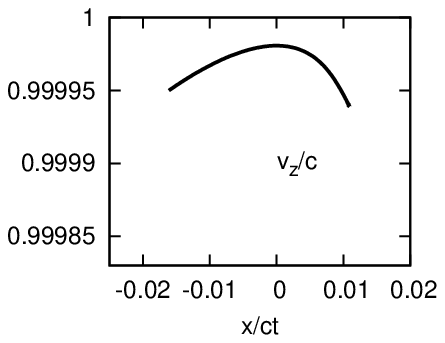}
\includegraphics[width=43mm,angle=0]{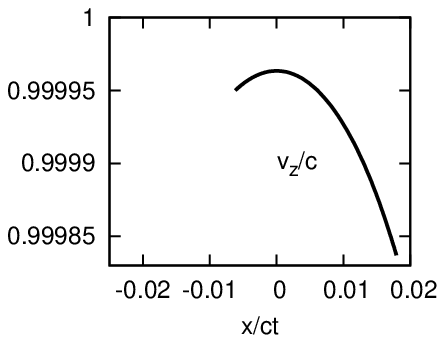}
\includegraphics[width=43mm,angle=0]{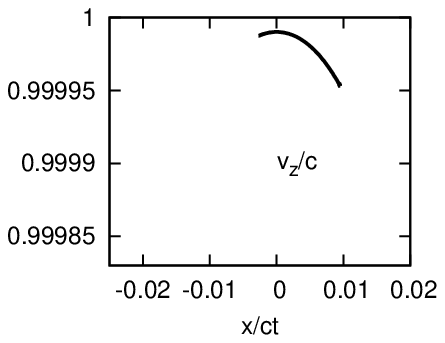}
\\
\includegraphics[width=43mm,angle=0]{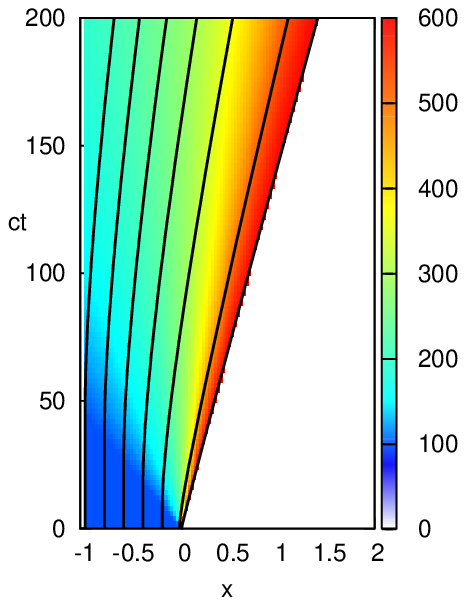}
\includegraphics[width=43mm,angle=0]{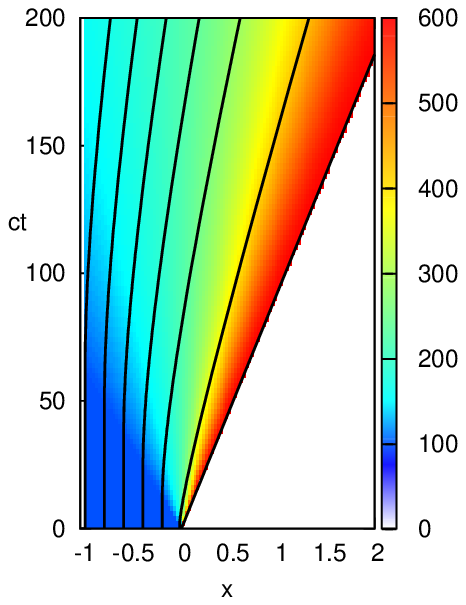}
\includegraphics[width=43mm,angle=0]{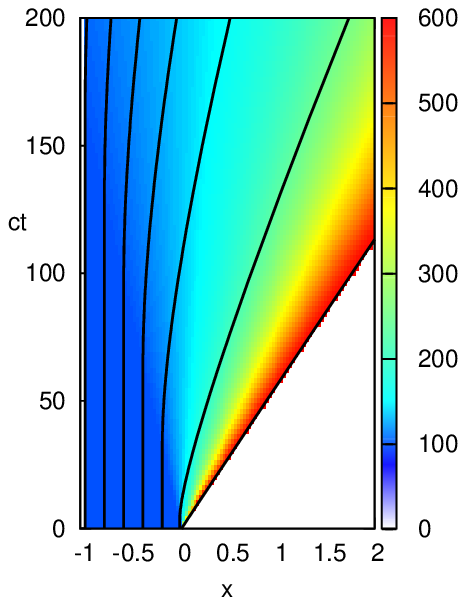}
\includegraphics[width=43mm,angle=0]{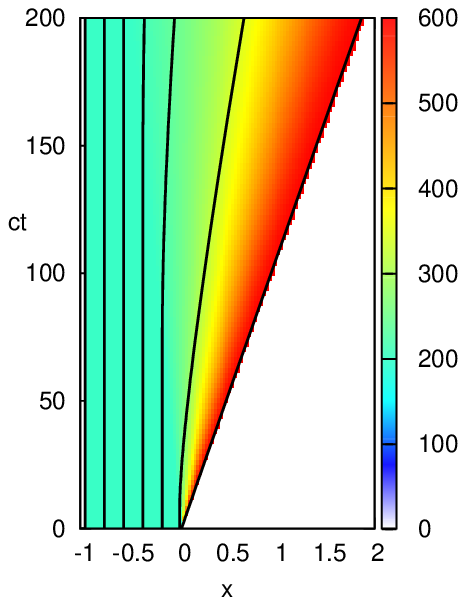}
\caption{Simple rarefaction wave solutions of relativistic MHD.
Each column corresponds to a particular Riemann problem. The parameters of the 
left state are given at the top of the column and the right state is always vacuum. 
In all cases $\mu_j=600$ and the polytropic index is $\gamma=4/3$. 
The last row shows the Minkowski diagrams for the solutions.  Here the colour image   
represents the distribution of the Lorentz factor (color) and the contours show 
the worldlines of fluid parcels initially located at 
$x_i=-1, -0.8, -0.6, -0.4, -0.2, -0.02, 0$.
The head of the rarefaction wave can be seen on these plots as the location 
where the Lorentz factor starts to grow and the worldlines start to bend to 
the right.  
}
\label{raref_gammacont}
\end{figure*}

In our jet simulations we deal with cold ($\cs=0$), super--fast-magnetosonic 
($\Gamma_j \gg \sigma_j^{1/2}$), relativistic ($v_j\simeq c$) jets. 
In this limit we have  
\begin{eqnarray}
  h_g = 1+\sigma,\quad h=1,\quad \frac{\cf^2}{c^2} =
  \frac{\sigma}{1+\sigma}, \quad  v_z\simeq c\, .
\nonumber \end{eqnarray}
Equation~(\ref{uz}) then gives  
\begin{eqnarray}
\Gamma=\frac{\mu_j}{1+\sigma} \,,
\nonumber \end{eqnarray}
whereas equation~(\ref{Ib}) yields 
\begin{eqnarray}
   \sigma = \sigma_j \frac{\rho}{\rho_j} \,.
\nonumber \end{eqnarray}
Combining the last two results we find that  
\begin{equation}
\Gamma=\frac{\mu_j}{1+\sigma_j \rho/\rho_j} \,.
\label{Lor1}
\end{equation}
Moreover, in this limit the integral in equation~(\ref{Ip}) 
assumes a simple analytic form. Specifically,
when $\cs=0$ and $v_z \approx c$ we have 
$u_z \approx \Gamma =\mu_j / (1+\sigma)$ and thus
\begin{eqnarray}
  X(\rho) = \frac{\sigma^{1/2}}{(1+\sigma)^2+\mu_j^2} 
\left[ (1+\sigma)^3 +\mu_j^2 \right]^{1/2} \,.
\nonumber \end{eqnarray}
In the super--fast-magnetosonic regime $\Gamma_j \gg \sigma_j^{1/2}$ and 
$ \mu_j^2 = (1+\sigma_j)^2 \Gamma_j^2 \gg (1+\sigma_j)^3$.
This makes it possible to simplify the expression for $X(\rho)$ 
even further,
\begin{eqnarray}
  X(\rho) = \frac{\sigma^{1/2}}{\mu_j} 
= \frac{\sigma_j^{1/2}}{\mu_j} \left(\frac{\rho}{\rho_j}\right)^{1/2},
\nonumber \end{eqnarray}
which leads to 
\begin{eqnarray}
\frac{1}{2}\ln\fracb{c+v_x}{c-v_x} = \frac{2\sigma_j^{1/2}}{\mu_j} 
   \left[1-\left(\frac{\rho}{\rho_j}\right)^{1/2} \right]. 
\nonumber \end{eqnarray}
This shows that $v_x\ll c$, which enables us to approximate the left
side of this equation as $v_x/c$ and write 
\begin{equation}
\frac{v_x}{c}
=\frac{1}{\Gamma_j} \frac{2 \sigma_j^{1/2} }{1+\sigma_j} 
\left[1-\left(\frac{\rho}{\rho_j}\right)^{1/2}\right]\, .  
\label{exp_speed}
\end{equation}
For the jet problem this implies that the jet opening half-angle $\theta_v$  
increases only by the amount 
\begin{equation}
\Delta\vvartheta =\frac{1}{\Gamma_j} \frac{2 \sigma_j^{1/2} }{1+\sigma_j} 
 < \frac{1}{\Gamma_j} \, .
\label{Dtheta}
\end{equation}
These considerations imply that the product $\Gamma \vvartheta$ can
increase to values that are significantly greater than 1 and that this
occurs mainly on account of the increase in $\Gamma$. This is indeed the
behaviour seen in our jet simulations.

\subsection{Hydrodynamic versus magnetic mechanisms}
\label{hdvsmhdsection}

The rarefaction acceleration process has been seen in numerical
simulations of both unmagnetized and magnetized flows and a number of
conclusions have been drawn about the role of the magnetic field
\citep{M08}. As both cases may have applications in astrophysics, we are
motivated to extend our analysis and investigate the role of the
magnetic field in the rarefaction mechanism a bit further. In
particular, it is interesting to see if the presence of strong magnetic
fields can lead to some observationally identifiable features. To check
on ths we have derived self-similar solutions of the Riemann problem
described in Section~\ref{case2} for four different left states: a cold
MHD flow, a hot MHD flow and two purely hydrodynamic (HD) flows with
different values of $\Gamma_j$. In all of these cases the magnitude of
the energy flux per unit rest-mass energy flux $\mu_j$ is kept the same.
The Lorentz factor near the boundary with the vacuum approaches this
value, independently of the other characteristics of the rarefaction
wave (see equation~\ref{boundary-G}).  However, the spatial distribution
of the Lorentz factor near the right boundary 
turns out to be quite sensitive to the magnetization of the left
state.

Fig.~\ref{raref_gammacont} shows the results of the numerical
integrations for an ideal gas with polytropic index $\gamma=4/3$.  If we
compare the properties of the cold MHD flow (first column) with those of
the HD flow (third column) we see that the acceleration is much stronger
in the cold MHD case. In the first place this is due to the fact that,
in the MHD case, the acceleration occurs over a larger volume of the
jet. This is simply a reflection of the difference in the speeds of the
rarefaction wave's head -- in the HD case it is determined by the sound
speed $\cs<c/\sqrt{3}$, whereas in the MHD case it is determined by the
fast-magnetosonic speed $\cf$, which can approach the speed of light.

A second reason for the difference between the MHD and HD cases has to
do with the dependence of the (generalized)
specific enthalpy on the mass density. 
Whereas $b^2/4\pi\rho c^2$ is proportional to the mass density
$\rho$, $w/\rho c^2$ is proportional to
$\rho^{\gamma-1}=\rho^{1/3}$.  As the density drops across the
rarefaction wave, the magnetic energy declines much faster than the
thermal energy, which explains the relative inefficiency
of the hydrodynamic case. For example, in the HD flow shown in the
third column of Fig.~\ref{raref_gammacont}, the density is
$\rho=\rho_j/125$ at $x/ct=0.01$, but the value of $w/\rho c^2$ is still
2 (meaning that the Lorentz factor is only a half of its maximum value,
$\Gamma=\mu_j/2$). In contrast, in the cold MHD case $b^2/4\pi\rho c^2$
is only $1/25$ at the same value of the density, meaning that $\Gamma$
is already $\approx \mu_j$.  Note that the time dependence of the
density of a particular fluid parcel is different in the two cases. As
shown in Appendix~\ref{append_approx}, in the MHD case the density drops
with time as $\rho \propto T^{-2/3}$, whereas in the HD case it declines
faster ($\rho \propto T^{-6/7}$) due to the higher speed of the tail of
the rarefaction.  Nevertheless, the (generalized) specific
enthalpy in the MHD and HD cases is proportional
to $\rho \propto T^{-2/3}$ and $\rho^{1/3} \propto T^{-2/7}$,
respectively, implying that the increase of the Lorentz factor is much
stronger in the former case.

\subsection{The role of the bunching function}
\label{bunching}

The above interpretation of the rarefaction-acceleration mechanism is
consistent with the general analysis of the acceleration of cold,
steady-state jets presented in \citet{KVK09}, where it was shown that
the bulk flow acceleration is intimately related to the shape of the
field (or stream) lines through the `bunching function'
\begin{equation}
\quad {\cal S}=\frac{\pi r^2 B_{\rm p}}{\Psi} \,,
\label{BF}
\end{equation}
where $r$ is the cylindrical distance from the rotation axis, 
$B_{\rm p}$ is the amplitude of the poloidal magnetic field and $\Psi=\int
\bmath{B}_p \! \cdot \! d \bmath{S}$ is the magnetic flux
function. 
In the super--fast-magnetosonic regime the Lorentz factor increases when ${\cal S}$ 
decreases. Specifically, one can show that  
\begin{equation}
\left(\frac{v^2}{c^2} -\frac{\mu-\Gamma}{\Gamma^3}\right) \frac{d(\Gamma
v)}{d \ell} = -\frac{2}{mc} \frac{d{\cal S}}{d \ell}\, ,
\label{mom}
\end{equation}
where $\ell$ is the distance measured along the poloidal magnetic field
lines and the effective rest-mass
$m$ is constant on magnetic flux surfaces (see Appendix~\ref{appB}). 
The coefficient in front of the derivative on the left hand side of this 
equation vanishes at the fast-magnetosonic critical surface 
which, in the highly relativistic limit, implies 
$\Gamma \approx \mu^{1/3}$.
Thus, in the super--fast-magnetosonic regime, acceleration corresponds to
a decrease in ${\cal S}$, whereas in the sub--fast-magnetosonic  
regime it corresponds to an increase in the bunching function. 
This is analogous to the transonic hydrodynamic flow in 
a de Laval nozzle,
with $1/{\cal S}$ playing the role of the nozzle cross section.

Equation (\ref{BF}) shows that, for ${\cal S}$ to decrease, $B_{\rm p}$
should decrease faster than $r^{-2}$, i.e. the separation between
neighbouring magnetic flux surfaces should increase with distance 
faster than $r$, their cylindrical radius. 
In confined flows, such as model B1 and the other cases
studied in \citet{KVK09}, this is realised through the stronger collimation
of the inner flux surfaces relative to the outer ones. For this
reason, the acceleration mechanism at work in such flows can be dubbed
the {\it collimation mechanism}.

\begin{figure}
\centerline{\includegraphics[width=80mm]{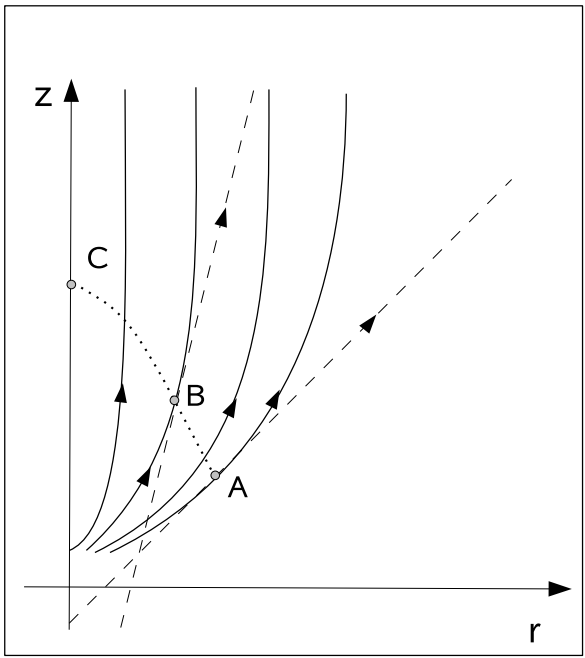}}
\caption{Straightening of the jet magnetic field lines after the channel
opens up.  In this sketch the solid lines show the magnetic field of an
externally confined jet, the dashed lines show the radial magnetic field
lines of a ballistic jet, and the dotted line shows the location of the
rarefaction wave front.  The magnetic field lines of a jet that
becomes free after encountering  a sudden widening of the confining wall
(point A) are represented by the solid lines upstream of the rarefaction
front and by the dashed lines downstream of it (where we assume for simplicity
that the field lines straighten out right after crossing the front).  The
figure shows two such field lines, which cross the rarefaction front at
points A and B, respectively. As one moves along the jet from A to B the
inner line is still parabolic whereas the outer one is already
straight. Thus, the separation between these lines increases faster than $r$
and the amplitude of poloidal magnetic field decreases faster than $r^{-2}$.
} 
\label{ill}
\end{figure}

\begin{figure}
\includegraphics[width=90mm,angle=-90]{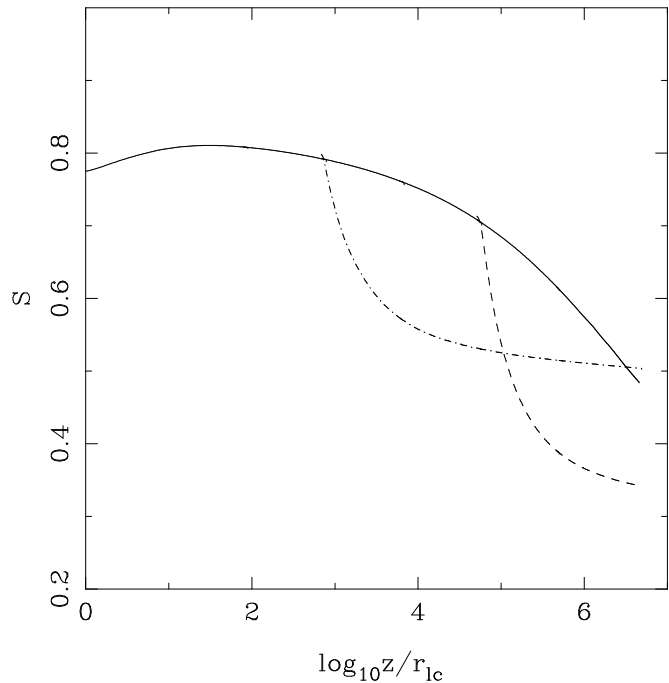}
\caption{Evolution of the bunching function ${\cal S}$ along the magnetic 
surface with $\Psi=15$ for models B1 (solid line), B1a (dash-dotted line) 
and B1b (dashed line). 
}
\label{S}
\end{figure}

The acceleration mechanism that operates in models B1a and B1b during
the transition to the ballistic regime is different from the collimation
mechanism in that it involves a rarefaction wave, and we therefore dub
it the {\it rarefaction mechanism}. In this process, the rarefaction
wave that is launched at the jet boundary at the point where the channel
widens reaches the jet axis much further downstream. Therefore, the
outer field lines straighten much closer to the source than the inner
ones (see Fig.~\ref{ill}). The net effect is again that $B_{\rm p}$
decreases faster than $r^{-2}$. Inspection of models B1a and B1b shows
that the magnetic bunching function indeed decreases along the magnetic
field lines (see Fig.~\ref{S}) and that in both of these models the flow
becomes super--fast-magnetosonic well upstream of the channel widening
point.\footnote{The sub--fast-magnetosonic regime is applicable in
model E of \cite{KVK09}, for which it was found that, in contrast with
the super--fast-magnetosonic regime considered in this paper, a widening
of the jet boundary does not lead to a significant acceleration of the
flow.}

\section{Discussion and conclusions}
\label{conclusion}

Our numerical simulations confirm the discovery by \citet{T09b} of the
additional fast acceleration of relativistic jets during the transition
from a confined to an unconfined regime.  We have analyzed the
acceleration mechanism and concluded that it is related to the sideways
expansion of the jet that is triggered by the 
elimination of support from the confining wall. 
This expansion does not lead to a significant increase of the jet
opening angle because of the high Mach number that the flow has already
attained by the time it reaches the location where the channel opens
up. The induced rarefaction wave is nevertheless strong enough to lead
to the conversion of a large fraction of the remaining magnetic energy
into kinetic energy of bulk motion.

The rarefaction-acceleration mechanism and its potential relevance to
relativistic astrophysical flows were previously considered by
\citet{AR06} and \citet{M08} for unmagnetized and magnetized flows,
respectively.  
As discussed in Section~\ref{hdvsmhdsection}, the slow expansion of the 
left front of the rarefaction and the weak dependence of the enthalpy
on the density in a hydrodynamic outflow, as compared to the 
faster expansion and stronger dependence of the magnetic energy 
on the density in MHD cases, is evidently behind the finding 
of \citet{M08} that this acceleration mechanism is 
more efficient in MHD outflows than in purely hydrodynamic ones.

Although this mechanism operates even in Newtonian
flows, its role there is not that important. This is because a high
fast-magnetosonic Mach number in the Newtonian regime implies that
most of the magnetic energy has already been converted into kinetic
energy of bulk motion, so the effect of any additional acceleration is
rather insignificant even if it involves a full conversion of the
remaining magnetic energy. The situation is different in the
relativistic case since the flow can remain magnetically dominated
even in the highly super--fast-magnetosonic regime.

Note in this connection that the gradual increase of the Lorentz
factor of an unmagnetized long/soft GRB outflow as it first emerges
from underneath the surface of the associated progenitor star
\citep[e.g.][]{MLB07} has a different origin than the rarefaction
acceleration considered here. In the unmagnetized case, a `cocoon' of
moderately relativistic, shocked jet and stellar material that formed
while the outflow was confined within the star is the first component
to be revealed. The second component to emerge is the faster shocked
jet material that occupies the region between the jet head and the
reverse shock that was induced by the interaction of the outflow with
the stellar envelope. Finally, the unshocked, high-velocity jet itself
comes out. 
The behaviour of the latter component after it leaves the star is
initially influenced by the presence of the `shielding' cocoon gas.
Whether significant rarefaction acceleration could occur under these 
circumstances even in a jet that reaches the stellar surface with 
a measurable fraction of its thermal energy not yet converted into 
kinetic energy remains to be investigated. 

We have shown that when the rarefaction wave propagates inside the jet
it produces a region where the separation between neighbouring
poloidal magnetic surfaces increases faster than their cylindrical radius. This
is a common feature of magnetic acceleration in the
super--fast-magnetosonic regime and is also a characteristic of the
magnetic acceleration process of externally confined flows by the
collimation mechanism. However, there are also significant differences
between these two mechanisms. Most importantly, acceleration by the
collimation mechanism can be sustained over many decades in distance and
produces asymptotic flows with $\sigma\simeq 1$, whereas the rarefaction
mechanism continues only for as long as it takes for the rarefaction
wave to cross the jet (of the order of the current jet length in our numerical
models), and subsequently the flow enters a phase of ballistic expansion
wherein the magnetic acceleration is so slow that it can be
ignored. 

We emphasize that the present study has involved axisymmetric simulations. 
A fully three-dimensional (3D) treatment is needed to address the issue
of stability, although we note that the results to date \citep[e.g.][]{MB09}
indicate that magnetically accelerated relativistic jets basically
remain stable as they propagate away from the source. It is also not yet
entirely clear whether the acceleration/collimation properties of
3D outflows remain the same as in the axisymmetric case, but, at least on
the basis of Newtonian simulations \citep[e.g.][]{Moll09}, it appears that
the acceleration efficiency remains roughly the same.

The most effective magnetic acceleration of steady flows could be achieved
via a combination of both mechanisms -- first the collimation mechanism
produces a flow with $\sigma\simeq 1$, and then the rarefaction
mechanism provides additional acceleration, resulting in a
particle-dominated ($\sigma \la 1$) flow.  Model B1b is 
an illustrative example of such a combination (see
Fig.~\ref{sigma}), and it is interesting to examine whether it could
in principle be realized in long/soft GRB sources. 
If the jet originates from a rapidly rotating black hole
then the light cylinder radius is of the order of a few gravitational radii,
$r_{\rm lc}\simeq 2\!\times\!10^6\,$cm for a $3\, M_\odot$ black hole. If the
confining medium is the stellar envelope, the collimation mechanism
operates up to $\sim 10^5r_{\rm lc}$. A quick inspection of Fig.~\ref{sigma}
shows that this corresponds to the channel widening point in our model
B1b. Thus, it may indeed be possible to accelerate long/soft GRB flows to the
high Lorentz factors indicated by observations, 
with $\sigma$ decreasing to values $\la 1$ (as happens in this example
on scales $z\simeq10^{12}\,$cm). 
However, it seems unlikely that the low magnetization ($\sigma \ll
1$) required for effective 
dissipation in the internal-shocks model for the prompt $\gamma$-ray
emission can be attained in this scenario, which suggests that
magnetic energy dissipation might need to be invoked to explain this
emission \citep[e.g.][]{LB03}.
If the initial magnetization $\sigma_0$ is very high, well above $\sim 10^3$, the jet 
will still be magnetically dominated when it enters the ballistic regime.

As pointed out by \citet{T09b}, the rarefaction-acceleration mechanism 
can in principle give rise to magnetically accelerated outflows that
satisfy $\Gamma\theta_{\rm j}\gg1$. In the 
standard model of uniform GRB jets, this condition allows a late' (on a
time scale $\ga 1\;$ day)
panchromatic jet break in the afterglow light curve. However, as the
asymptotic structure of magnetically accelerated  jets is far from being
uniform, further investigation is required to fully establish this result. 
Furthermore, in some of the GRB sources
detected by {\it Swift}\/ there have been indications of an early (on a
time scale of $\ga 1\,$hr) jet break \citep[e.g.][]{Pan07,Kam09},
corresponding to $\Gamma\theta_{\rm j}\la 1$ and $\theta_{\rm
j}\la 1^\circ$, as predicted by pure collimation-acceleration
models \citep{KVK09}. If these inferences are corroborated by further
observations, the reasons why certain long/soft GRB sources show no evidence of
rarefaction acceleration will need to be clarified. It is, however, also
conceivable that the acceleration of jets that exhibit a late break in
the afterglow lightcurve is predominantly thermal, since in this case
the value of $\Gamma\theta_{\rm j}$ is not limited as in the magnetic
acceleration scenario. One may, however, be able to distinguish between
the thermal and magnetic mechanisms -- at least in some cases -- based on other
observational diagnostics, such as the appearance of a photospheric
emission component \citep[e.g.][]{Ryd10}. This issue could
potentially be further illuminated by studies of short/hard GRB
afterglows. Since the outflows in the latter sources evidently do not
propagate through a stellar envelope, the rarefaction-acceleration
mechanism would not operate in this case (or at least not in the same
way as in long/soft GRB  sources). If short/hard GRB outflows are accelerated
predominantly by magnetic stresses, this suggests that there should be no
late-time breaks in their afterglow lightcurves (or at least fewer than
in the case of long/soft GRBs). While the data on short/hard GRB
afterglows are still sparse \citep[e.g.][]{Nak07}, this prediction could
in principle be tested as more such afterglows are observed.

\section*{Acknowledgments}
SSK was supported by the STFC grant
``A Rolling Programme of Astrophysical Research at Leeds''.
NV acknowledges partial support by the Special Account 
for Research Grants of the National and Kapodistrian University of Athens.
AK was partially supported by a NASA Astrophysics Theory Program grant.


\appendix 

\section{Interpretation of equation~(\ref{EQ-CHAR1})}
\label{Eq20}

The derivation of the characteristics equations that yield the phase
speeds of relativistic MHD waves can be found in a number of references
\citep[e.g.][]{AP87,A89,K99}. For the fast-magnetosonic waves the result is 
\begin{equation}
\begin{array}{l}
\Gamma^4(\lambda-v_x)^4(1-\cf^2) + (1-\lambda^2)\ \times  \\
\quad \\
\left[
  (\cs^2/w_g)(b^x-\lambda b^0)^2-\Gamma^2(\lambda-v_x)^2 \cf^2
   \right]=0\,,
\end{array}
\label{i1}   
\end{equation}   
where $w_g=w+b^2/4\pi$ is the generalized enthalpy and $b^\nu$ is the 
magnetic field 4-vector (see equations~29 and~A20 in \citealp{K99}). 
In order to have compact equations we employ in this appendix geometric units 
in which $c=1$. The components of 
$b^\nu$ can be obtained given the magnetic field and flow velocity
3-vectors via 
\begin{equation} 
  b^0=\Gamma B_i v^i, \qquad b^i=(B^i-\Gamma v^i b^0)/\Gamma\,.  
\label{i2}
\end{equation}  
In the problem under consideration $b^{\tilde{0}}=0$ and $b^{\tilde{x}}=0$. 
Moreover, in the frame comoving
with the fluid along the $x$ axis $\tilde{v}_x=0$ and equation~(\ref{i1})
yields
\begin{equation}
  \tilde{\lambda}^2=\cf^2 \left[\tilde{\Gamma}^2+\cf^2(1-\tilde{\Gamma}^2)\right]^{-1} .
\label{i3}
\end{equation}  
Here we use a tilde to indicate quantities in the comoving frame
(comoving only in the $x$ direction). Using the equations of Lorentz
transformation we find
\begin{equation}
\tilde{\Gamma} = \bar{\Gamma}\Gamma(1 -\bar{v} v_x)\, , 
\label{i4}
\end{equation}  
where $v_i$ and $\Gamma$ are the flow parameters as measured in the frame of 
the initial discontinuity and where $\bar{\Gamma}$ and $\bar{v}$ describe
the motion  of the comoving frame. Since $v_x=\bar{v}$, this equation yields   
\begin{equation}
\tilde{\Gamma}^2=\frac{\Gamma^2}{\bar{\Gamma^2}} = 
\frac{1-v_x^2}{1-v_x^2-v_z^2} = 1+u_z^2\,,
\label{i5}
\end{equation}  
where $u_z=\Gamma v_z$.
Substituting this result into equation~(\ref{i3}), we obtain 
\begin{equation}
\tilde{\lambda}^2=\cf^2\left[1+u_z^2(1-\cf^2) \right]^{-1}.
\label{i6}
\end{equation}
Thus, the right-hand side of equation~(\ref{EQ-CHAR1}) is indeed the
fast-magnetosonic wave speed as measured in a frame that comoves with
the fluid along the $x$ axis.

\section{The rarefaction wave in the ultra-relativistic limit}
\label{append_approx} 

The Riemann problem considered in Section~\ref{case2} is described by
equations~(\ref{Is})--(\ref{Ip}) and (\ref{EQ-CHAR1}).  Here we add that
one can follow the wordline of a fluid parcel initially located at $x_i
(<0)$, by integrating the equation
\begin{equation} 
\frac{d\xi}{d\ln t}=v_x-\xi\, , 
\end{equation} 
which follows from the definition $\xi\equiv x/t$.
The dimensionless time $T\equiv ct/(-x_i)$, for times $T>T_i\equiv
-1/\xi_{\rm head}$ (i.e. after the head of the rarefaction wave has
already passed $x_i$), is given by
\begin{equation}
T = T_i \exp\int^{\rho_j}_\rho \frac{-\frac{d\xi}{d\rho}}{v_x-\xi} d\rho
\,. 
\label{teq}
\end{equation}

\subsection{The ultra-relativistic cold MHD limit}
\label{MHD}
The expressions for the Lorentz factor $\Gamma$ and the expansion speed
$v_x$ as functions of the density $\rho$ in the cold $(\cs=0)$,
ultra-relativistic $(v_z\approx c)$ limit have already been 
derived in Section~\ref{case2} and are given by equations~(\ref{Lor1})
and~(\ref{exp_speed}), respectively. Using the same simplifications, we
can find the density and the Lorentz factor as functions of $\xi=x/t$. 
Equation~(\ref{EQ-CHAR1}) can be writen as
\begin{equation}\label{xoverctsimple}
\frac{x}{ct}=\frac{1}{\mu_j} 
\left[2\sigma_j^{1/2} -3 \left(\sigma_j \frac{\rho}{\rho_j}\right)^{1/2} 
-\left(\sigma_j \frac{\rho}{\rho_j}\right)^{3/2} \right] \,,
\end{equation}
implying $(x/ct)_{\rm min}=-\sigma_j^{1/2}/\Gamma_j$ for the head
and $(x/ct)_{\rm max}=2\sigma_j^{1/2}/\mu_j$ for the tail of the
rarefaction wave. The Lorentz factor is
$\Gamma=\mu_j / \left(1+\sigma_j \rho/\rho_j \right)$, with 
\begin{equation}\label{rhosimple}
\rho=\frac{4\rho_j}{\sigma_j} \sinh^2 \left[
\frac{1}{3}{\rm arcsinh} \left(
\sigma_j^{1/2}-\frac{\mu_j}{2}\frac{x}{ct}
\right)\right] \,,
\end{equation}
found by inverting equation~(\ref{xoverctsimple}).
We can also express the Lorentz factor as a function of 
the dimensionless time $T$, since equation~(\ref{teq}) implies
\begin{equation}
\rho=\rho_j \sigma_j^{-1/3}\Gamma_j^{2/3} T^{-2/3} \,.
\end{equation}
The resulting expression is
\begin{equation}
\Gamma=\frac{\mu_j}{1+\sigma_j^{2/3}\Gamma_j^{2/3} T^{-2/3}} \,.
\label{gam_analMHD}
\end{equation}

\subsection{The ultra-relativistic HD limit}
\label{HD}
Similar approximations in the purely hydrodynamic limit ($\sigma_j=0$) imply
\begin{eqnarray}
\Gamma=\frac{\mu_j}{1+\left(h_j-1\right)\varrho^{\gamma-1}} 
\label{gammaHD} \,, \\
\frac{v_x}{c} =\frac{
{\cal I} \left[\left(2-\gamma\right)\left(h_j-1\right)\right]
-{\cal I} \left[\left(2-\gamma\right)\left(h_j-1\right)\varrho^{\gamma-1}\right]}
{\mu_j \left(\gamma-1\right)^{1/2}\left(2-\gamma\right)^{1/2}} 
\,, \nonumber \\
{\cal I} \left[\zeta\right] \equiv \zeta^{1/2} \left(1+\zeta\right)^{1/2}
+\ln \left[\zeta^{1/2}+\left(1+\zeta\right)^{1/2}\right] \,,
\\
\frac{x}{ct}=\frac{v_x}{c} 
-\frac{1}{\Gamma} \left[
\frac{\left(\gamma-1\right)\left(h_j-1\right)\varrho^{\gamma-1}}
{1+\left(2-\gamma\right)\left(h_j-1\right)\varrho^{\gamma-1}}
\right]^{1/2} \,, \\
T =\Gamma_j \varrho^{-(\gamma+1)/2} \left[   
\frac{1+\left(2-\gamma\right)\left(h_j-1\right)\varrho^{\gamma-1}}
{\left(\gamma-1\right)\left(h_j-1\right)}
\right]^{1/2}
\,,
\label{timeHD}
\end{eqnarray}
where $\varrho \equiv \rho/\rho_j$. (Note that, for $\gamma \rightarrow
2$ and $h_j\rightarrow 1+\sigma_j$,
we recover the cold MHD case considered in Section~\ref{HD}.)

For $\gamma=4/3$ and $\Gamma$ significantly larger than $0.4 \mu_j$, 
equation~(\ref{timeHD}) simplifies to
\begin{equation}
\varrho=3^{3/7} \left(h_j-1\right)^{-3/7} \Gamma_j^{6/7} T^{-6/7} \,.
\end{equation}
Subtituting this result into equation~(\ref{gammaHD}), we infer the
dependence of the Lorentz factor on $T$,
\begin{equation}
\Gamma=\frac{\mu_j}{1+
3^{1/7} \left(h_j-1\right)^{6/7} \Gamma_j^{2/7} T^{-2/7}} \,.
\label{gam_analHD}
\end{equation}

Comparing the results~(\ref{gam_analMHD}) and~(\ref{gam_analHD}), we see
that the rarefaction-acceleration process is faster in the MHD case than
in a purely hydrdynamic flow. This is part of the reason for why this
mechanism is more efficient in MHD jets, as discussed in
Section~\ref{hdvsmhdsection}. 

\section{The bunching function and magnetic acceleration}
\label{appB}

At distances from the central source where the flow can be considered
cold and the azimuthal velocity small, the component of the momentum
equation along the motion can be written as

\begin{equation}\label{momentum-V_p}
\frac{\Gamma \rho}{2}
\left(1+\Gamma^2 \frac{v^2}{c^2} \right) \frac{d v^2}{d \ell}=
-\frac{B_{\phi}}{4 \pi r} \frac{d \left(r B_{\phi} \right)}{d \ell}\ ,
\end{equation}
where $\rho$ is the rest-mass density, $r$ is the cylindrical coordinate, 
$B_\phi$ is the azimuthal component of the magnetic field, 
$\ell$ is the arclength along a poloidal field (or stream) line,
and derivatives with respect to $\ell$ are taken along a given field line
(see e.g. equation~20 in \citealp{VK03a}).

We can further simplify equation~(\ref{momentum-V_p}) by using two
integrals of motion, the mass flux per unit magnetic flux $(\massint)$
and the angular velocity of magnetic field lines $(\Omega)$ (see e.g.
Section~2.1 in \citealp{KVK09}).  Using their expressions in the limit
of a small azimuthal velocity, we have
\begin{equation}\label{psi_a}
\massint=\frac{\rho \Gamma v}{B_p} \,, 
\quad
\Omega=-\frac{v}{r} \frac{B_\phi}{B_p}\,.
\end{equation}
Equation~(\ref{momentum-V_p}) can be written in the form of a
momentum equation of a point particle in a potential field:
\begin{equation}
\label{momentum1}
m \frac{d (\Gamma v)}{dt} = - \frac{d{\cal V}}{d \ell}\ .
\end{equation}
The effective rest mass, $m$, is inversely proportional to the magnetization,
\begin{equation}
\label{mass}
m=\frac{8\pi^2\massint c}{\Psi \Omega^2} 
\,,
\end{equation}
and the effective potential energy is related to the bunching function,
\begin{equation}
\label{potential}
{\cal V} = \frac{2\cal S}{v/c}\ .
\end{equation}
The corresponding energy equation can be written as
\begin{equation}\label{integral}
m \Gamma c^2 + {\cal V} = m \mu c^2
\,, \end{equation}
where the integral of motion $\mu$ represents the 
total energy flux per unit rest-mass energy flux.

The bunching function ${\cal S}$ is directly connected to the geometry 
of the flow. The cross sectional area between two neighbouring 
flux surfaces $\Psi$ and $\Psi+\delta \Psi$ is $\delta \Psi/B_p$.
Thus, the function ${\cal S}$ is proportional to $r^2$ over this area
and decreases whenever the flow expands in such a way that
this area increases faster than $r^2$
(\citealp{V04}; see also Section~5.1 in \citealp{KVK09}).
The evolution of ${\cal S}$ is determined by the
transfield component of the momentum equation; it
depends on the external pressure that confines the jet
(or, equivalently, on the shape of the `wall' that defines its outer
boundary). 

Since the potential depends not only on ${\cal S}$ but also on 
the velocity, it is useful to separate out
the effect of geometry in the equation of motion. 
Combining equations~(\ref{momentum1}), (\ref{potential})
and~(\ref{integral}) we can write
\begin{equation}
\label{momentum2}
\left(\frac{v^2}{c^2} -\frac{\mu-\Gamma}{\Gamma^3}\right) \frac{d(\Gamma
v)}{d \ell} = -\frac{2}{mc} \frac{d{\cal S}}{d \ell}\ .
\end{equation}
The resulting critical point corresponds to the fast magnetosonic surface,
where $d S / d \ell =0 $ and $\Gamma \approx \mu^{1/3}$
(in the relativistic regime where $1 \ll \Gamma\ll \Gamma^3$).
Acceleration in the super-fast (sub-fast) magnetosonic regime corresponds to
decreasing (increasing) ${\cal S}$, respectively. 

If at some point the curvature of the jet boundary suddently increases,
the adjustment of the magnetic field from the old to the new curvature
corresponds to a fast expansion and decrease of the bunching function.
This is precisely the effect of the {\it rarefaction wave}\/ analysed in
Section~\ref{interpretation}.  As an example, suppose that the initial
shape is parabolic, $z=c_1 r^{c_2}$, and the final shape is conical, $z=z_0+r/\tan
\vartheta_{\rm tr}$, as shown in Fig.~\ref{ill}. The variables $z_0$ and
$\vartheta_{\rm tr}$ are functions of $\Psi$ and can be found from a
smooth matching of the magnetic field along the transition surface
$z_{\rm tr}=z_{\rm tr}(r_{\rm tr})$ (dotted line in
Fig.~\ref{ill}). Downstream from this surface the bunching function
${\cal S}$ declines as
\begin{equation} 
{\cal S}=\frac{{\cal S}_{\rm tr}+\Delta {\cal S}}
{1+\displaystyle\frac{\Delta {\cal S}}{{\cal S}_{\rm tr}}
\displaystyle\frac{r_{\rm tr}}{r}} \,, \qquad \Delta {\cal S}=-{\cal
S}_{\rm tr} \frac{\sin^2 \vartheta_{\rm tr}}{r_{\rm tr}}
\frac{dz_0}{d\vartheta_{\rm tr}} 
\label{eq_for_S} 
\end{equation} 
(see equation~35 and the related discussion in \citealp{V04} for further
details). The
appearance of the variable $z_0(\Psi)$ in the equation describing the
conical boundary ($z=z_0+r/\tan \vartheta_{\rm tr}$) is crucial for enablig the
additional acceleration (by making $\Delta {\cal S}<0$): this is
elaborated in the discussion of the difference between `type I conical'
(in which $z_0=0$) and `type Ia conical' (in which $z_0 \ne 0$)
shapes in \cite{V04}; similar categories exist for
parabolic shapes as well (type I, Ia, II and IIa).  Although the adopted
channel shape in model B1b is a simple cone whose vertex is located at the
origin, the field lines clearly have a `type I conical' shape, i.e.
their projection crosses the $r=0$ axis at $z_0<0$ with $dz_0/d\Psi>0$
(see the right panel of Fig.~\ref{lorentz} and Fig.~\ref{ill}).

Equation~(\ref{eq_for_S}) fits well the curves seen in Fig.~\ref{S}.
In one decade or so in cylindrical distance downstream from
the transition radius $r_{\rm tr}$, the bunching function
drops from ${\cal S}_{\rm tr}$ to ${\cal S}_{\rm tr} +\Delta {\cal S}$.  
The difference $\Delta {\cal S}$ ($<0)$ corresponds to a difference 
$\Delta {\cal V}\approx 2\Delta {\cal S}$ in the potential energy and thus to
an acceleration $\Delta \Gamma = -2\Delta {\cal S} / mc^2$.

\end{document}